\documentclass[ijoc,nonblindrev]{informs3} 

\OneAndAHalfSpacedXII 

\usepackage{natbib}
\usepackage{subfiles}
\usepackage{algorithmic}
\usepackage{graphicx}
\usepackage{textcomp}
\usepackage{multicol}

\usepackage{mathtools}
\usepackage{amsmath,amssymb}
\usepackage{amsfonts}
\usepackage{braket}

\usepackage{breqn}

\usepackage{dsfont}
\usepackage{lineno}
\usepackage[T1]{fontenc}

\bibpunct[, ]{(}{)}{,}{a}{}{,}%
\def\bibfont{\small}%
\TheoremsNumberedThrough     

\EquationsNumberedThrough    

\begin{document}


\RUNAUTHOR{Mohammadbagherpoor, et.al.}

\RUNTITLE{Gate-Scheduling QAP Using Quantum Computers}

\TITLE{Exploring Airline Gate-Scheduling Optimization Using Quantum Computers}

\ARTICLEAUTHORS{%
\AUTHOR{Hamed Mohammadbagherpoor}
\AFF{NC State University, \EMAIL{hmohamm2@ncsu.edu}}
\AUTHOR{Patrick Dreher}
\AFF{NC State University, \EMAIL{padreher@ncsu.edu}}
\AUTHOR{Mohannad Ibrahim }
\AFF{NC State University, \EMAIL{mmibrah2@ncsu.edu}}
\AUTHOR{Young-Hyun Oh }
\AFF{IBM Corporation, \EMAIL{ohy@us.ibm.com}}
\AUTHOR{James Hall}
\AFF{NC State University, \EMAIL{jlhall4@ncsu.edu}}
\AUTHOR{Richard E Stone}
\AFF{Delta Air Lines, \EMAIL{richard.stone@delta.com}}
\AUTHOR{Mirela Stojkovic}
\AFF{Delta Air Lines, \EMAIL{mirela.stojkovic@delta.com}}

} 

\ABSTRACT{
%
This paper investigates the application of quantum computing technology to airline gate-scheduling quadratic assignment problems (QAP). We explore the quantum computing hardware architecture and software environment required for porting classical versions of these type of problems to quantum computers. We discuss the variational quantum eigensolver and the inclusion of space-efficient graph coloring to the Quadratic Unconstrained Binary Optimization (QUBO). These enhanced quantum computing algorithms are tested with an 8 gate and 24 flight test case using both the IBM quantum computing simulator and a 27 qubit superconducting transmon IBM quantum computing hardware platform.
}


\maketitle

%


\section{Introduction}

One of the many complex problems facing the airline industry today is how to optimize large combinations of cargo and passenger traffic among the planes, gates, personnel and the air traffic flows originating or terminating at an airport.  Finding solutions to this complex combinatorial optimization problem is important for airlines in order to be able to control costs and minimize inefficiencies (\cite{Stollenwerk2019}).

A mathematical framework that can be used to compute optimal solutions for this type of problem is known as the quadratic assignment problem (QAP). This is a difficult combinational optimization problem of the NP-hard class (\cite{Sahni1976}). 

A widely cited application of a QAP is the assignment of facilities to locations such that each facility is assigned to only one location. Given that facility versus location assignment costs are known, the problem is finding a minimum cost allocation of facilities to locations taking the costs as a sum of all distance and flow products. A constrained QAP is a natural framework for assigning flights to gates under a variety of constraints and objectives. Generally, such QAP type problems of size n $>$ 30 cannot be solved optimally on digital computers in reasonable time.  For example, see (\cite{Loiola2007}), or (\cite{GPBFD}). Improvements through approximations, heuristics and special structures (\cite{Kim2017}) can all be implemented in order to get reasonable solutions for some, but not all, purposes and time frames. The question remains if one can do better than reasonable or do more than some.

One expectation of a viable quantum computer is the capability to find an optimal solution to large quadratic assignment problems. Delta Air Lines has partnered with North Carolina State University in an effort to explore how quantum computing may be able to address such types of flight-gate scheduling optimizations. The goal will be to understand what are the potential capabilities for a quantum computer to be able to optimally solve large quadratic assignment problems that are so computationally prohibitive they could never be computed using classical digital computers. 

In Section~\ref{sec:Classical-Computing-Approach} we present a combinatorial optimization model (CO) model that is a natural QAP approach to gate scheduling.  In Section~\ref{sec:QC-Approach} we describe the quantum computing hardware architecture and software environment used when porting this type of model to a quantum computer.  In Section~\ref{sec:results} we discuss the simulations and experimental results and analysis and comparison between the digital and quantum computing methods.  Finally in Section~\ref{sec:Conclusion} there is a discussion of conclusions that can be inferred from the work and a look toward future steps for improving this algorithm as quantum computing hardware platforms become more robust.




\section{Digital Computing Implementation of Gate Scheduling Optimization Algorithms}
\label{sec:Classical-Computing-Approach}

The natural base formulation of the gate-scheduling problem is a QAP where the decision is which gate to assign to each plane. The objective is naturally quadratic with the major component being the interaction of pairs of planes. For this paper we will use the passenger walking distance between gates given a predicted value of the number of passengers connecting between planes.  To model the plotting problem at a fixed point in time the following variables need to be specified. 

\begin{itemize}
    \item  $P$ = Set of planes to park at gates
    \item $G$ = Set of gates at which to park planes
    \item $v_{ij}$ = Expected number of passengers connecting between planes $p_i$ and $p_j$
    \item $d_{kl}$ = Walking distance between gates $g_k$ and $g_l$
    \item $x_{ik} = 1$ if plane $p_i$ is parked at gate $g_k$, $0$ otherwise 
\end{itemize}

The challenge is to minimize the objective function $\mathcal{J}$ with respect to the constraints that follow:

\begin{equation}
\mathcal{J}=\sum_{i,j \in P} \ \sum_{k,l\in G} {{x_{ik}}{x_{jl}}{v_{ij}}{d_{kl}}}\
\label{eq:1}
\end{equation}

such that 

\begin{equation}
\sum_{k\in G} {x_{ik}} = 1 \hspace{1em} \forall \ i \in P 
\label{eq:2}
\end{equation}

\begin{equation}
\sum_{i\in P} {x_{ik}} = 1  \hspace{1em} \forall \ k \in G    
\label{eq:3}
\end{equation}

where $x_{ik} \in \{0,1\} \hspace{1em} \forall \ i \in P \ and \ \forall \ k \in G$ \\
 
Modifications to this initial generic formulation need to be made in order to arrive at a specific useful model. As an example, we would want to replace the constraint Eq.(\ref{eq:3}) with the following.
\begin{equation}
\sum_{i \in P} \ {x_{ik}} \ \leq 1 \hspace{1em} \forall \ k \in G\
\label{eq:4}
\end{equation}
The constraints in Eq.(\ref{eq:2}) and Eq.(\ref{eq:4}) mathematically state that a plane needs to be assigned a gate, but a gate does not need to be assigned a plane. 

This set of constraints should be modified further.
\begin{equation}
\sum_{i \in P_t} \ {x_{ik}} \ \leq 1 \hspace{1em} \forall \ k \in G \ \text{and} \ \forall \ t \in T
\label{eq:5}
\end{equation}
Here, $P_t$ is the set of planes requiring a gate at time $t$ in the time period $T$ under consideration  However, using all of $T$ would create redundant constraints. Fortunately, there are efficient ways to find a subset of $T$ which eliminates the redundancy without weakening this set of constraints. \\

There is a long list of modifications that should be considered for a useful model. Here is a small sampling from this list.

\begin{description}
\item[$\bullet$ \textbf{Gate Restrictions}] For various reasons, not every gate can accommodate every plane.\
\item[$\bullet$ \textbf{Gate Closings}] Not all gates can be used all the time. \
\item[$\bullet$ \textbf{Contentions}] Even if two particular planes can individually be assigned to two particular gates, proximity may preclude both assignments being made at the same time. \
\item[$\bullet$ \textbf{Alleyway Congestion}] Gates using the same alleyway should avoid planes departing
close in time. \
\item[$\bullet$ \textbf{Adjacent Gates}] Gates close in proximity should avoid planes departing close in time. \\
\end{description}

Solving a realistic version of this gate plotting model for a large and busy airport would be impractical by normal methods. An interesting aspect of quantum computing is that the underlying QAP itself seems to be a more natural problem to solve. What one avoids in classical computing is instead what one uses in quantum computing. Exploring the option for utilizing a quantum computer to optimize these gate-scheduling problems may be a viable path forward in the future.




\section{Quantum Computing Implementation of Gate Scheduling Optimization Algorithms}
\label{sec:QC-Approach}

Calculating a complete solution of a plane-gate scheduling constrained QAP problem for a very large and busy airport is impractical with digital computing technologies. A quantum computer may have the potential to circumvent some of these difficulties.

Exploring the possibility of implementing a gate scheduling optimization algorithm for a quantum computing system requires a re-design of the digital version.  This includes both developing a working understanding of both the quantum computing hardware architecture and software environment so that the digital version of this QAP algorithm can be successfully re-structured and migrated to a quantum computing hardware system.

\subsection{Quantum Computing Hardware Architectures}
\label{QC_hardware}

Quantum computers are based on a fundamentally different premise for computation when compared to a digital computer. The digital computer operates on the mathematics of a $base_{2}$ system, whereas the fundamental premise of computation for a quantum computer is based on something called a qubit. Unlike a classical bit that only has two states (sometimes designated as "On" or "Off" or "0" or "1"), a quantum bit (qubit) can be represented by many different weighted, linear combinations of "0" and "1". The rules for quantum computation are based on the physics of quantum mechanics and concepts of superposition and quantum entanglement rather than the electrical engineering design of digital logic circuits. (\cite{Nielsen2009})

Planning computations on a quantum computer and verifying proper operation of the algorithms running on the machine are also totally different from digital computations.  Unlike digital computations where it is possible to query a system's state at different points during the computation, it is impossible to observe the intermediate results for a quantum computation without destroying the quantum state. A system cannot be disturbed with any type of intermediate measurement until the completion of the calculation, otherwise the entire computation on the quantum computer will automatically be terminated in whatever state it was at when the measurement was attempted. There is no possibility of re-starting the computation from the point where an intermediate measurement was taken and continuing with that computation or of re-running a computation from the beginning in a deterministic way.  Upon completion of the computation, the idea of an "$answer$" also has a very different interpretation. Whereas a digital computer will give a deterministic answer, a quantum computer will provide a histogram representing a spectrum of measurements with their relative probabilities.

Today there are several different types of quantum computing hardware architectures that have been developed. These include an analog-based platform based on Quantum Annealing, various superconducting transmon devices based on non-linear Josephson junction circuits, and trapped ion devices.  In this paper, the IBM superconducting, transmon-based quantum computing hardware platforms were used for the computations.  We selected the \textit{IBMQ Toronto}, which is a 27-qubit quantum computing processor.

\begin{figure}[!t]
  \begin{center}
  \includegraphics[width=4.5in]{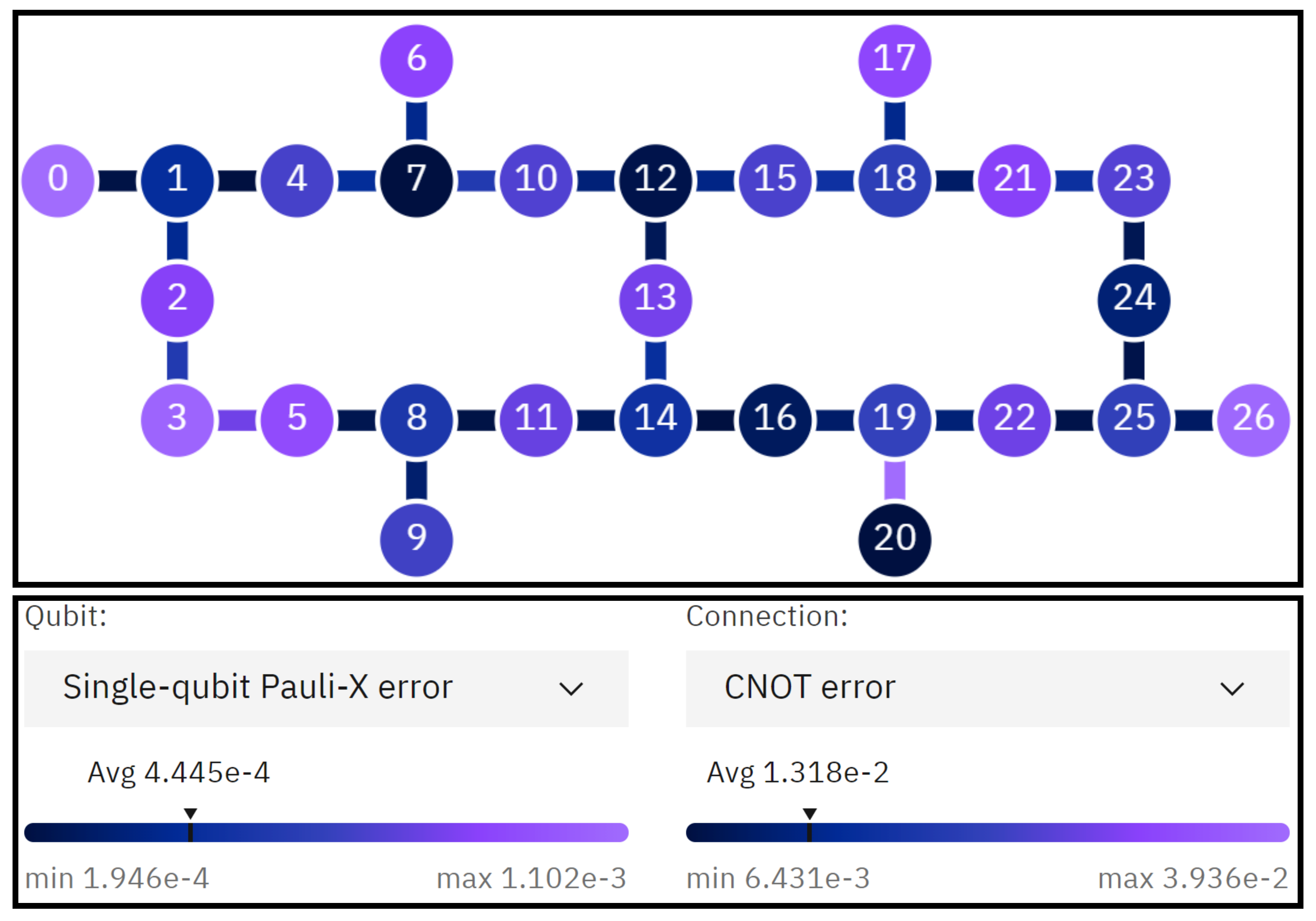}\\
  \caption{IBM published information  (IBM Quantum: (https://quantum-computing.ibm.com/) ) on the qubit layout on the IBMQ_Toronto quantum computing hardware processor.  The figure on the top is the actual connectivity of the qubits.  The figure on the bottom-left is a color coded "heat map" showing the individual 1 qubit error rates and on the bottom-right is the direct 2 qubit connection error rates on Toronto.}
  \label{IBMQ_Toronto}
  \end{center}
\vspace{-1em}
\end{figure}

Noise is an additional consideration when running on a quantum computing hardware platform.  Today's quantum computers are best described as Noisy Intermediate Scale Quantum (NISQ) machines.  Such types of noise present in these machines today adversely impacts the stability of the quantum computations.  Noise essentially destroys the coherence of the quantum state and limits the size of the application that can be run on these machines.  Mitigation of this noise and development of error correction methodologies that will extend the quantum computer's computational capabilities are both active area of research.  Fig.~\ref{IBMQ_Toronto} shows the general architecture and qubit connectivity of the machine and a graphical "heat map" showing the level of 1 qubit and 2 qubit errors for this processor. 

A critical component in the overall planning for the quantum computation is choosing the qubits for the computations on a quantum computing hardware architecture with the best profile for low error rates on that platform.  Quantum computing programs that run on today's NISQ platforms carefully place the quantum codes for that algorithm on the subset of qubits and connections with the best coherence and smallest error parameters in order to deliver a better execution of the program. Some additional properties of the Toronto processor and a brief discussion of error rates and their impact on quantum calculations are presented in the Appendix.

\subsection{Software Environments}
\subsubsection{Variational Quantum Eigensolver (VQE)}
\label{sec:algo1}

Quantum computers have the potential to solve specific mathematically structured classes of problems exponentially faster than classical computers. However,
developing the methodologies and software for achieving this goal will take research and development effort and may require new technologies to overcome today's NISQ computer limits. Because both the quantum hardware and software limitations will continue to be limiting factors in the near future, researchers have been investigating hybrid algorithms that can be run efficiently using a combination of both quantum and classical components. Eigenvalues, eigenvectors, and optimization problems play an important role in the development of such hybrid algorithms.

\cite{ref7} presented the Variational Quantum Eigensolver (VQE) algorithm as an improved method for finding the quantum state corresponding to the minimum energy of a Hamiltonian by applying a quantum/classical hybrid variational technique. In the VQE approach the Hamiltonian matrix $H$ can be derived by the interaction spins or electronic systems in physical systems. $H$ can be calculated to find the eigenvectors $\ket{\psi_i(t)}$ and the corresponding eigenvalues $\lambda_i$ of the Hamiltonian matrix by converting an optimization problem to an adiabatic quantum computation.  Using their technique, both classical and quantum computers are used together to iterate on an algorithm to converge intermediate results toward the correct solution.

The main goal of VQE is to find the desired eigenvectors and eigenvalues by optimizing a parameterized quantum program.  The lowest eigenvalue is of primary interest in the VQE approach since it is the optimum solution of the problem, same as the lowest energy is the solution to a Hamiltonian in physical systems.  VQE can be used to find the solution for Eq. (\ref{VQE}),
\begin{equation}\label{VQE}
E(\theta)=\underset{\theta} min\bra{\psi(\theta)}H\ket{\psi(\theta)}.
\end{equation}
Because $\ket{\psi(\theta)}$ is the normalized solution for Eq. (\ref{VQE}), it can be the minimum eigenvalue of $H$. $H = \sum_{i} \beta_i H_i$ where $H_i$ are the tensor products of Pauli operators and $\beta_i$ are the real coefficients.
By considering this reformulation for $H$ matrix, Eq.~(\ref{VQE}) can be represented as,
\begin{equation}\label{VQE2}
E(\theta) = \underset{\theta} min\bra{\psi(\theta)}H\ket{\psi(\theta)} = \underset{\theta} min \sum_{i} \beta_i \bra{\psi(\theta)}H_i\ket{\psi(\theta)}.
\end{equation}

\begin{figure}[!t]
  \begin{center}
  \includegraphics[width=4.5in]{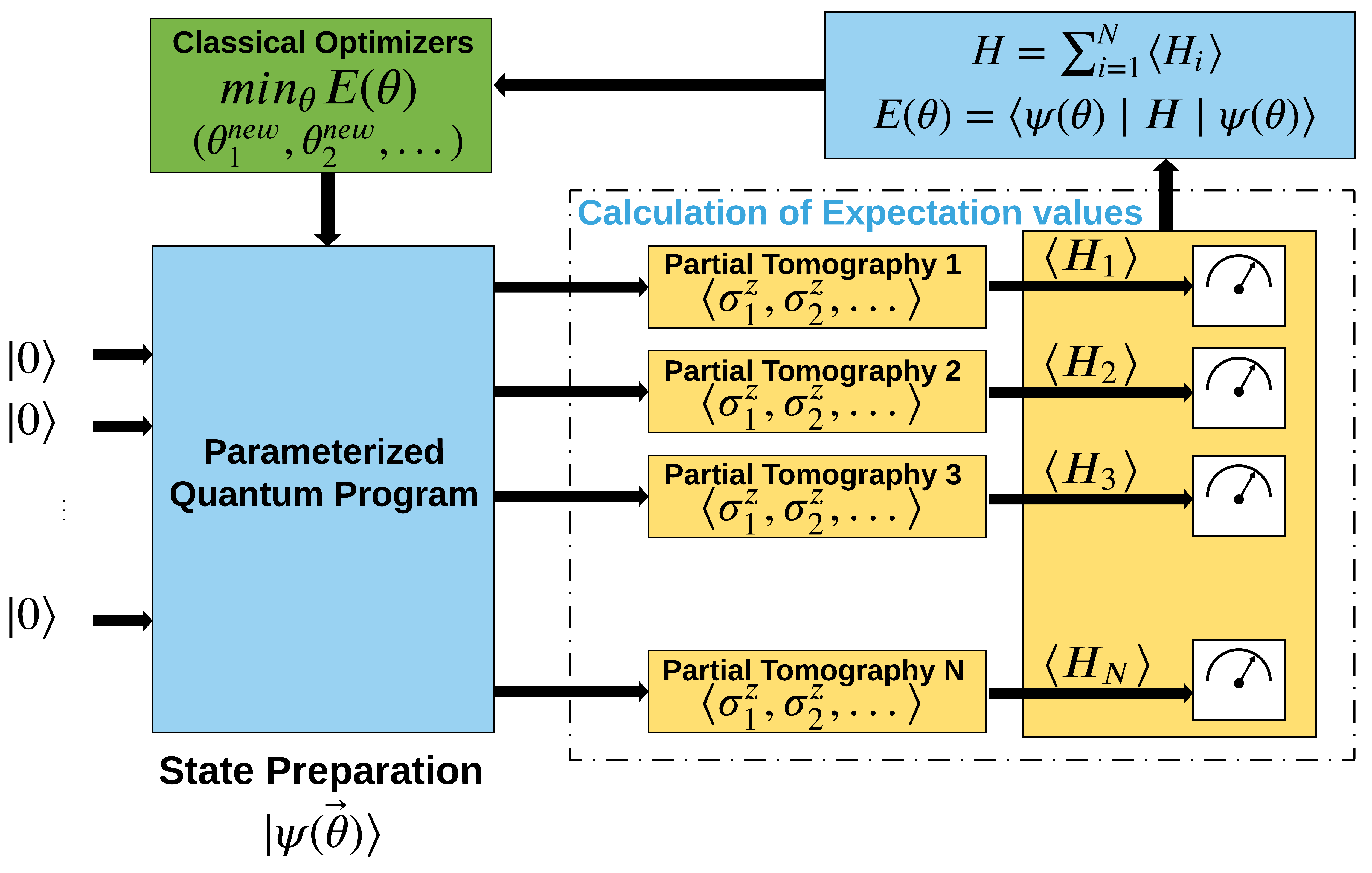}\\
  \caption{The schematic diagram of the variational quantum eigensolver (VQE) algorithm
}
  \label{VQE_generalview}
  \end{center}
\vspace{-1em}
\end{figure}
Finding the minimum eigenvalue can be an NP-hard problem as the size of the Hamiltonian matrix increases. To address this problem, different quantum techniques have been developed~(\cite{ref22}). Quantum phase estimation (QPE) is one of the main techniques that can be applied to a unitary quantum circuit derived from a problem to find its minimum eigenvalue, that is the solution of the Eq.~(\ref{VQE}). However, practical resource constraints with today's NISQ hardware limit the implementation of this algorithm. Hence, it is not possible to find a complete solution to this problem using near-term noisy quantum computers.

Alternatively, researchers have been developing some hybrid classical-quantum algorithms (e.g., VQE (\cite{ref7}) and QAOA (\cite{ref8, Wang2018AnIT, Guerreschi2019})) to solve a problem by finding the minimum eigenvalue of the Hamiltonian Matrix $H$ in a recursive optimization process using both NISQ quantum computers and classical computers. Fig.~\ref{VQE_generalview} illustrates the components of the VQE algorithm. In this hybrid system, all of the qubits in the quantum computer are initialized to the zero state. A parameterized quantum program (i.e., state preparation) is applied to the zero states and the expectation value of the corresponding Hamiltonian ($\bra{\psi(\theta)}H_i\ket{\psi(\theta)}$) is calculated.  A classical optimizer is used to update the parameters and find an optimum values for $\vec{\theta}$ through the classical optimization process~(\cite{ref8, ref7}). This process is repeated recursively until the hybrid system finds the minimum eigenvalue of the $H$ matrix~(\cite{ref23}) or some optimization limit are reached. The parameterized quantum circuit ($U(\theta)$) is known as an \textit{Ansatz}. Choosing an ansatz operator as a variational form plays an important role in the optimization process. In most cases, an ansatz should be able to efficiently run in quantum devices. 

Different classical optimizers can be used to solve the optimum values for $\theta$. These techniques perform various forms of gradient descent methods to apply iterative optimization algorithms to solve the problem~(\cite{ref16,ref17}).  

This paper considers two well-known classical optimizers. 
\begin{itemize}
    \item The constrained optimization by linear approximation (COBYLA) algorithm is a numerical optimization method that finds the min/max of the objective function without knowing the gradient value~(\cite{ref41}).
    \item The simultaneous perturbation stochastic approximation (SPSA) algorithm is used to optimize a system with multiple unknown parameters. This algorithm only uses the cost function measurements and is efficient in high-dimensional optimization problems. The gradient approximation using this technique requires only two measurements regardless of dimension of the problem which makes this algorithm to be a powerful technique to optimize a problem with multiple unknown parameters (\cite{SPSA1}-\cite{SPSA2}).
        
\end{itemize}
The implementation and results for each of the optimizers will be presented in section~\ref{sec:results}.

\subsubsection{Binary Optimization Problem and Quantum Hamiltonian}
\label{sec:binary-optimization}
Combinatorial optimization (CO) is a major class of optimization problems that has various applications in industry and research. In particular, CO for Graph Coloring has been extensively used in different application domains such as transportation management, social media, telecommunication, and resource scheduling. Most of these applications are NP-hard problems that are attractive but challenging for researchers to solve.


Three approaches can be considered to tackle CO: Exact Algorithms, Approximation Algorithms, and Heuristic algorithms. (\cite{Halim2017, Festa2014})

An exact algorithm, by definition, is one that is guaranteed to find a solution, if one exists. Therefore, unless P = NP, an exact algorithm cannot have worst-case polynomial running time for an NP-hard problem. Thus, in practice, an exact algorithm can take too much time to reach an optimal solution, though it might find good solutions in reasonable time.

An approximation algorithm is an exact algorithm to an approximate version of the problem. For example, an approximation algorithm might find the solution over a subset of the set of feasible solutions, or use an approximate version of the objective function.  Approximation algorithms are a trade off between speed and solution quality in the cases where known exact algorithms are too slow. Ideally, an approximation algorithm has a guarantee of how close it comes to optimal, e.g., the solution found will have an objective value at least 4/5 of the true maximum of the original problem. 

A heuristic algorithm, whether it works on the original problem or an approximate version, does not have a proven guarantee that it will find a solution in finite time for the problem on which it works.  Heuristic algorithms are usually faster and often simpler than other algorithms and,
so, represent another trade off between speed and solution quality.

Recently, Quadratic Unconstrained Binary Optimization (QUBO) has been introduced to cover a large variety of CO problems~(\cite{ref21,McGeoch2014, Kadowaki1998,Bertels2020}). The QUBO technique can find an optimal solution for many types of NP-hard problems such as graph partitioning, graph coloring, scheduling management, and register allocation.

A QUBO problem can be formulated by Eq. (\ref{QUBO}) as described in~(\cite{ref31}),
\begin{dmath}\label{QUBO}
 J = min \;\;\{X^T Q X + g^T X  + c\} = \sum_{i, j \in P} \sum_{k, l \in G} x_{i k} x_{j l} v_{i j} d_{k l}+\mathrm{P} \sum_{i}\left(1-\sum_{k} x_{i k}\right)^{2}+\mathrm{P} \sum_{(i, j)} \sum_{k} x_{i k} x_{j k}
\end{dmath}

\noindent where $Q$ is a $n\times n $ matrix with real numbers, $X$ is the state vector with element of $\{0,1\}^n$ and $P$ is a positive and large number. The goal of Eq. (\ref{QUBO}) is to find the minimum of the objective function which is the minimum eigenvalue of the matrix $H$. Eq. (\ref{QUBO}) can also be expressed by Eq. (\ref{QUBO_2}) as described in~(\cite{ref19}),
\begin{equation}\label{QUBO_2}
 \sum_{i,j} W_{ij} x_i (1-x_j) + \sum_{i} W_i x_i,
\end{equation}

\noindent where $W_i$ are the linear weights $Q_{ii}$ for $i=1,2,...,n$ and $W_{ij}$ are the quadratic weights $Q_{ij}$ for $i<j$. Here $x_i$ is a binary variable that can take $0$ or $1$ and as a result $x_i^2 = x_i$. In order to transfer Eq. (\ref{QUBO_2}) into an Ising model, the following substitution can be used as
\begin{equation}\label{QUBO_sub_1}
 x_i = \frac{Z_i + 1}{2},
\end{equation}

\noindent where $ x_i\in\{0,1\}$ and $Z_i \in \{-1,1\}$ for $i=0,...,n$.
By applying the substitution the Hamiltonian matrix would be (\cite{ref18, ref20}),
\begin{equation}\label{QUBO_sub}
\begin{multlined}
 \sum_{i,j} \frac{W_{ij}}{4} (1-Z_i) (1+Z_j) + \sum_{i} \frac{W_i}{2} (1-Z_i) \\=
 -\frac{1}{2} \sum_{i,j} W_{ij} Z_i Z_j + \sum_{i} W_{i} Z_i + c.
 \end{multlined}
\end{equation}

This Hamiltonian matrix will be a form of tensor products of Pauli $Z$ operators. Note that since the $Z$ is a diagonal matrix, the Hamiltonian matrix should be diagonal so that it can be used in a VQE optimization process. In order to solve the $k$-coloring problem using the standard QUBO optimization technique, $nk$ qubits are required. The required number of qubits is the main constraint of the standard QUBO optimization approach. \cite{Serra2021} has recently presented a quantum annealing (QA) technique to find near optimal solution for the QUBO optimization problems by preparing the ground state of the Ising spin Hamiltonian. In this paper we have considered a different type of quantum computer which is a universal Gate-based superconducting quantum computer and has capability to find the optimum solution to the combinatorial optimization problem. In addition, in section~\ref{sec:algo2} a new algorithm is introduced which solves the $k$-coloring optimization problem by applying $n\; log(k)$ number of qubits which exponentially reduces the required number of qubits compared to the standard QUBO technique.

\subsection{Space-Efficient Graph Coloring Quantum Computing algorithms}
\label{sec:algo2}
In section~\ref{sec:algo1}, we showed that the standard QUBO optimization algorithm requires $nk$ number of qubits which increases exponentially with the problem size. Therefore, it is infeasable to run a gate scheduling problem with a reasonable size on existing NISQ devices. 

In this section, a new space-efficient technique that maps a $k$-coloring problem to a lower number of qubits is explained. The number of colors is encoded to $m$ bits (total $2^m$ colors) and the goal is to assign a color labeled by $m$ bitstrings to each vertex, (\cite{ref42}-\cite{Space_Eff2}). As mentioned in the standard $k$-coloring approach, two neighboring nodes should not have the same color, therefore in the space-efficient technique a cost function should evaluate to nonzero if the neighboring nodes have the same color code. For a graph with adjacency matrix $A$, the graph coloring problem can be mapped to the ground state of the following Hamiltonian Eq.~(\ref{eq:Space_Efficient_H}).

\begin{equation}\label{eq:Space_Efficient_H}
\begin{multlined}
  H =\sum_{i,j} A_{ij} \sum_{a\in\{0,1\}^m}   \prod_{l=1}^{m} ( \mathds{1}+(-1)^{a_l} Z_{i,l})( \mathds{1}+(-1)^{a_l} Z_{j,l})
  \end{multlined}
\end{equation}

The ground state of the $n\; log(k)$-qubit Hamiltonian matrix is the solution to the graph coloring problem which corresponds to the proper bitstring assignment of the graph vertices. If the number of colors is not a power of $2$, i.e. $2^{m-1}<k<2^{m}$, $m$ number of bitstrings is considered to map the colors, however, the bitstrings in which $\sum_{i=1}^m 2^{k-i}b_i<k$ are allowed for vertex coloring. This means that new penalty terms should be added to Eq.~(\ref{eq:Space_Efficient_H}) to satisfy this limitation. For example in case $k=3$, two bitstrings $\{00,01,10,11\}$ should be considered to label the colors. Since only three colors are required to be assigned to the vertices, a term $\sum_{i=1}^n (1-Z_{i,1})(1-Z_{i,2})$ should be added to the Hamiltonian matrix to penalize the non-allowed bitstring which is $(b_1,b_2) = (1,1)$. By applying Eq. (\ref{eq:Space_Efficient_H}) and considering the new penalty term, the corresponding Hamiltonian matrix will be,

\begin{equation}\label{Space_Efficient_H2}
\begin{multlined}
  H =\sum_{i,j=1}^n A_{ij} ( \mathds{1}-Z_{i,1})( \mathds{1}-Z_{i,2})( \mathds{1}-Z_{j,1})( \mathds{1}-Z_{j,2})\\
\;\;\;\;\;\;\;\;\;\;\;\;\;\;\;\;\;+( \mathds{1}+Z_{i,1})( \mathds{1}+Z_{i,2})( \mathds{1}+Z_{j,1})( \mathds{1}+Z_{j,2})\\
\;\;\;\;\;\;\;\;\;\;\;\;\;\;\;\;\;+( \mathds{1}+Z_{i,1})( \mathds{1}-Z_{i,2})( \mathds{1}+Z_{j,1})( \mathds{1}-Z_{j,2})\\
\;\;\;\;\;\;\;\;\;\;\;\;\;\;\;\;\;+( \mathds{1}-Z_{i,1})( \mathds{1}+Z_{i,2})( \mathds{1}-Z_{j,1})( \mathds{1}+Z_{j,2})\\
`+( \mathds{1}-Z_{i,1})( \mathds{1}-Z_{i,2})\;\;\;\;\;\;\;\;\;\;\;\;\;\;\;\;\;\;\;\;\;\;\;\;\;\;\;\;\;\;\;\;\;\;\;\;\;\;\;\;\;\;\;\;\;\;
  \end{multlined}
\end{equation}

As seen in Eq. (\ref{Space_Efficient_H2}), the last term will be added since four colors has been considered for a problem that only requires three colors. This penalty term helps remove one of the binary coded colors from the list.



\section{Simulation and Experimental Results}
\label{sec:results}

This section discusses the simulation and experimental results for both the standard QUBO optimization technique and the space-efficient coloring scheme.  All quantum simulations and computations on both IBM's quantum simulator and quantum computing hardware platforms were accessed through the NC State University Quantum Computing Hub connection into the IBM Quantum Network.

The gate scheduling problem described in this paper was formulated and mapped as a graph coloring problem and tested on both a quantum simulator and an actual quantum computing hardware platform.  We show both simulation and experimental results of applying both graph coloring implementations discussed in Sections~\ref{sec:algo1} and \ref{sec:algo2} to find an optimal solution for an example gate allocation problem and compare the results by considering different criteria such as number of required qubits, run-time and circuit depth. 

\begin{figure}[!t]
  \begin{center}
  \includegraphics[width=3.5in]{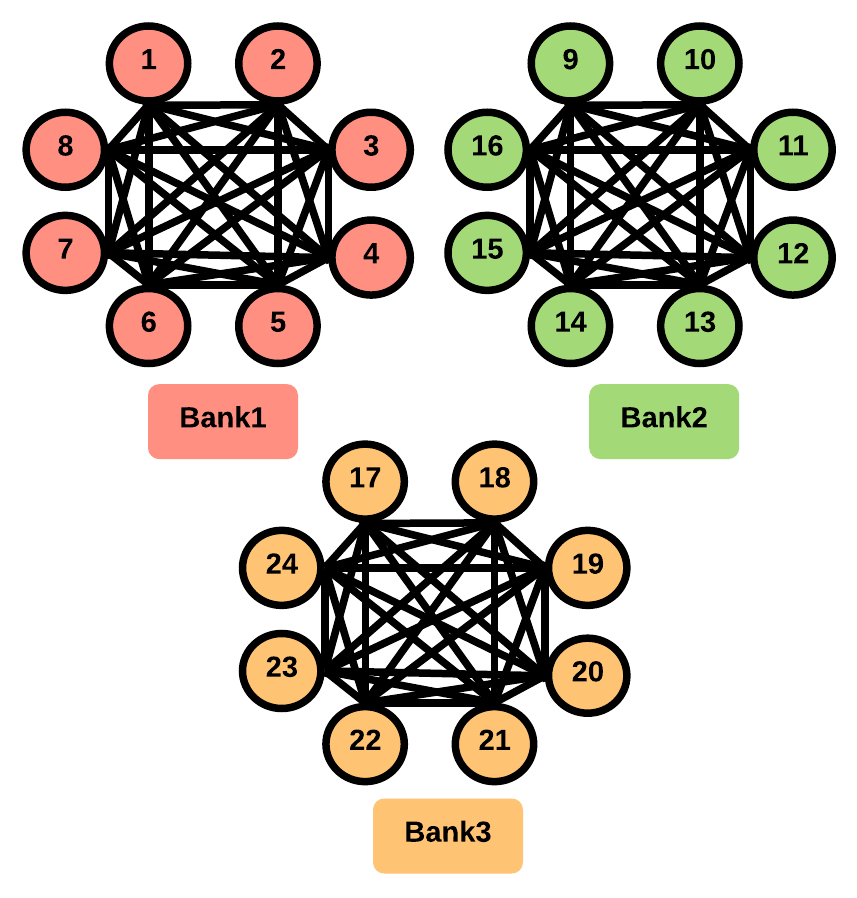}\\
  \caption{The corresponding derived graph based on the flights arrival and departure time interval.
}
  \label{Gate_Flight_Graph}
  \end{center}
\vspace{-1em}
\end{figure}

For our example test case, we consider a setup with eight gates and 24 flights. Fig.~\ref{Gate_Flight_Graph} shows the corresponding derived graph based on the flights arrival and departure time interval. The graph nodes correspond to the flights and an edge between two nodes shows the time overlap between the two flights. The goal here is to colorize the graph using quantum embedding approaches to minimize the problem objective function. By considering Eq.~(\ref{QUBO_sub}) and Eq.~(\ref{eq:Space_Efficient_H}) and converting the graph coloring problem to the corresponding Hamiltonian matrices, the problem was mapped onto the IBM Qiskit platform, (\cite{ref01}) and used to solve for the ground state of the Hamiltonian matrices.

A \textit{Noise-Aware Compilation} method from Qiskit was used for the circuit compilation and optimization. This method maps the circuit to match the hardware topology as discussed in Section \ref{QC_hardware} and also does light-weight optimizations by collapsing adjacent gates as appropriate. We notice a direct benefit from the efficient embedding in all aspects of the comparison. As both circuits run the same ansatz architecture, we see that the efficient embedding proves advantageous, as it uses fewer qubits. 

\begin{figure}[!ht]
  \begin{center}
  \includegraphics[width=.8\linewidth]{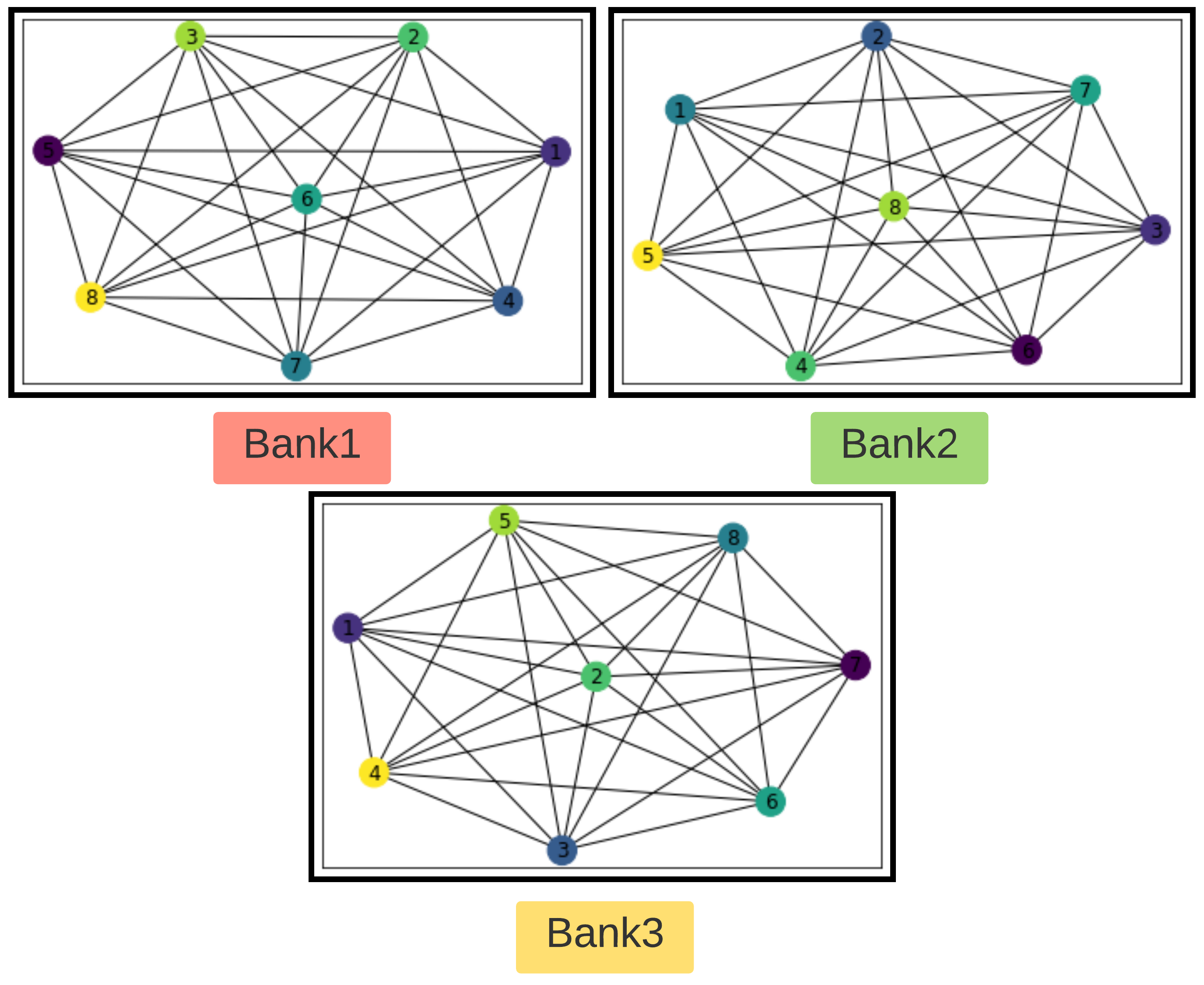}\\
  \caption{Quantum simulation results for $24$ flights and eight gates. The graph successfully colored by the applying the efficient embedding technique.
}
  \label{Gate_Flight_24flight}
  \end{center}
\vspace{-1em}
\end{figure}

\begin{figure}[!ht]
  \begin{center}
  \includegraphics[width=.8\linewidth]{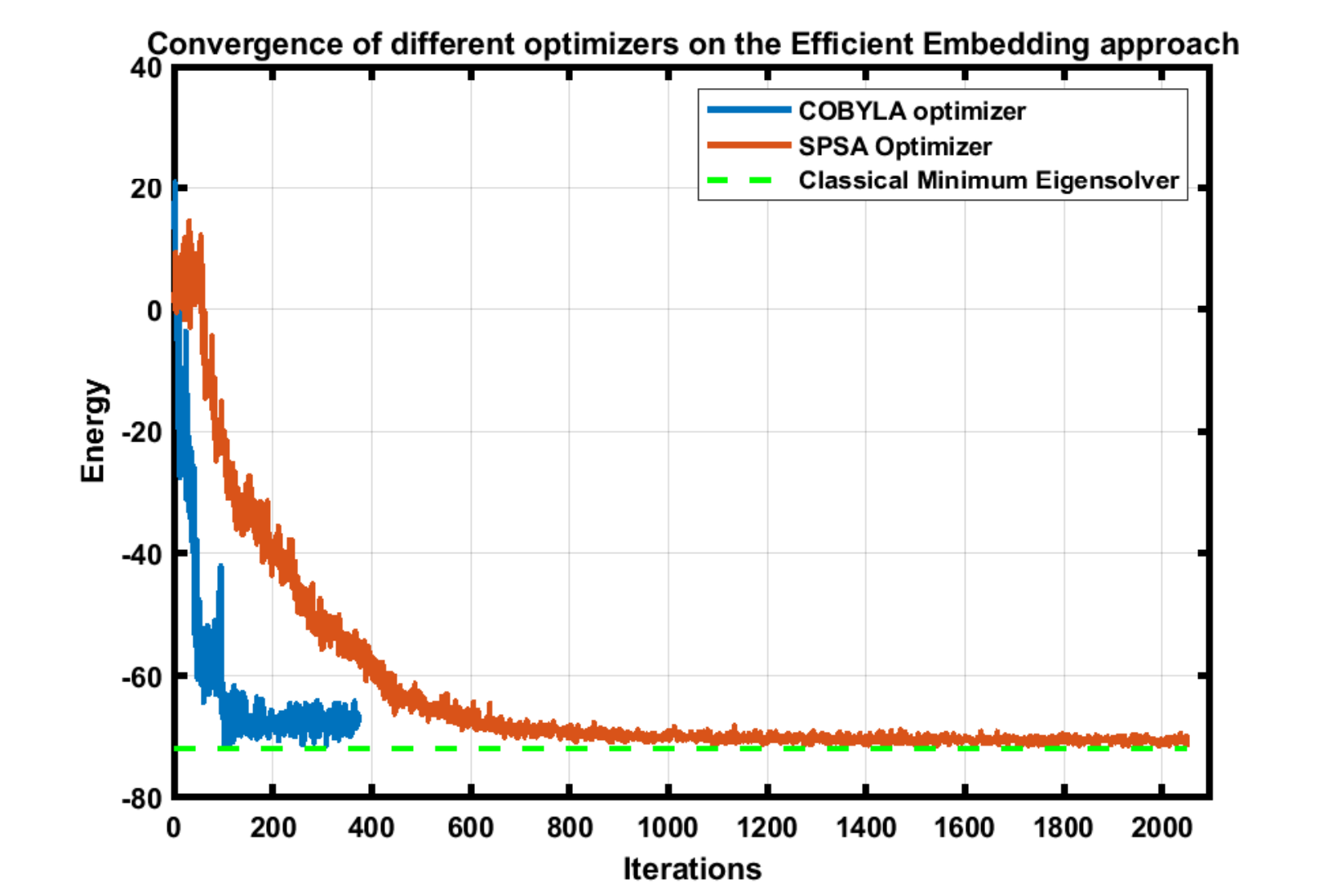}\\
  \caption{Convergence of different optimizers methods to update the parameters of quantum circuit. 
}
  \label{figure_Optimizers}
  \end{center}
\vspace{-1em}
\end{figure}

As discussed in the hardware architecture section (Section \ref{QC_hardware}) efficient embedding for mapping the problem to qubits on the hardware platforms focuses on the minimization of the usage of SWAP operations. Quantum algorithms formulated with fewer 2-qubit interactions will require fewer SWAPS to implement them on quantum platform. A design change requiring fewer 2-qubit operations results in a shorter circuit (program) depth with a more efficient embedding and hence a shorter run-time. As discussed in Section \ref{sec:algo2}, the efficient embedding scheme requires fewer qubits.  This shorter run-time allows larger gate-scheduling problem sizes to be implemented on the quantum platform. Simulation results were obtained using IBM's Aer simulator. Experimental results were obtained by running the circuits on IBM's Toronto 27 qubit quantum computing hardware platform. Fig. \ref{Gate_Flight_24flight} shows that the graph was successfully colored by applying the quantum space efficient embedding technique. As it can be seen the quantum machine is able to find the optimum solution for this problem. We investigate different classical optimizers, considering different well-known methods such as COBYLA and SPSA to optimize the parameters of the quantum circuits. The results has been shown in Fig. \ref{figure_Optimizers}. The optimizers have been evaluated under the same situation such as type of variational quantum circuit, depth of quantum circuit. The simulation results demonstrate that the optimizers converge to the correct value and are able to find an optimal solution for the flight gate allocation problem. It can be seen that the COBYLA has faster convergence rate compared to the SPSA algorithm. however, it was shown that the SPSA algorithm has better performance on today's NISQ machines to optimize the parameters of quantum variational circuits.

\begin{figure}[!ht]
  \begin{center}
  \includegraphics[width=\linewidth]{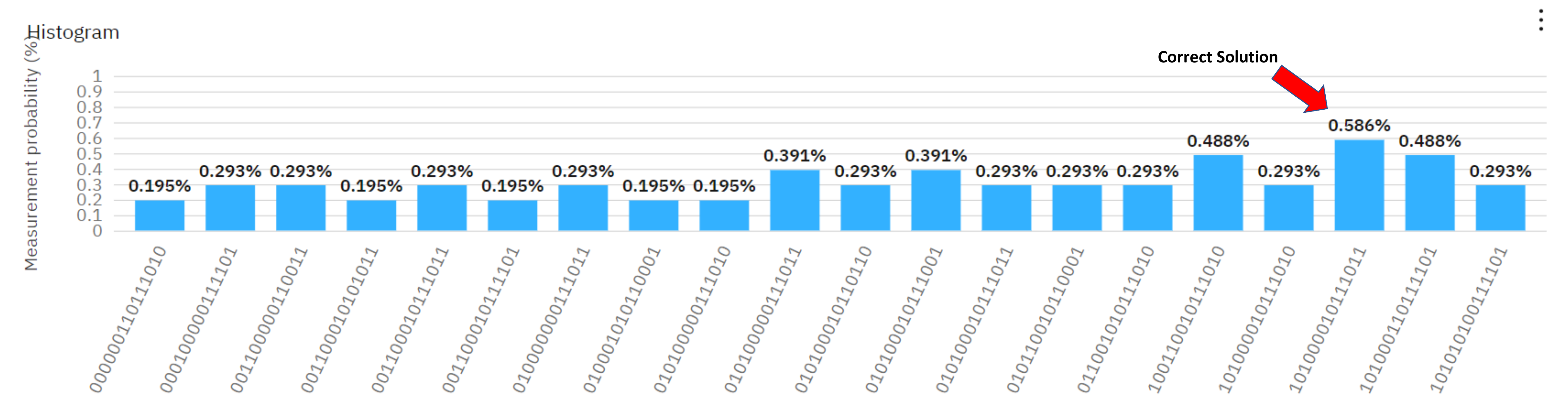}\\
  \caption{Hardware results with the measurement probabilities for a bank with eight flights and eight gates.
}
  \label{Gate_Flight_8_hardware}
  \end{center}
\vspace{-1em}
\end{figure}

Due to the required number of qubits and the software/hardware limitation, the previous case study (24 flights and 8 gates) cannot be solved by the standard graph coloring technique using today's quantum technologies. In order to evaluate and compare the presented quantum graph coloring techniques in terms of the number of qubits, program depth and run-time, a case study of flight gate scheduling problem with five flights has been considered. Both techniques were applied and the derived Hamiltonian matrices were run on the quantum machine. Table \ref{results_table} shows a comparison between the two implementations described in sections \ref{sec:algo1} and \ref{sec:algo2} in terms of the number of qubits, run-time and circuit depth. As it can be seen for the standard embedding, the required number of qubits exponentially increases as the size of the problem increases. The efficient embedding technique will reduce the number of qubits exponentially and requires smaller program depth which results in lower run-time on NISQ quantum machines to solve an optimization problem. Fig. \ref{Gate_Flight_8_hardware} shows the measurement probability of the possible solutions for this case study by running the corresponding Hamiltonian, derived by efficient embedding technique, on a noisy IBM quantum machine. The SPSA optimizer was utilized to optimize the circuit parameters. As can be seen the correct solution has the highest probability among the other possibilities and IBM-Q machine was capable of solving this optimization problems however, there is still challenges to tackle the optimization problems with relevant practical dimension due to the hardware limitation (noise), the required number of qubits and program depth.


\begin{table}[]
\caption{Comparison between the standard QUBO and the efficient embedding approaches}
\label{results_table}
\centering
\begin{tabular}{|l|c|c|}
\hline
                       & \textbf{Standard Embedding} & \textbf{Efficient Embedding} \\ \hline
\textbf{\# Qubits}     & $n \times k$: 25                          & $n \times log(k)$: 15                           \\ \hline
\textbf{Circuit Depth} & 29                           & 19                            \\ \hline
\textbf{Run-time (sec)}      & 5569.84                           & 395.48                           \\ \hline
\end{tabular}

\end{table}

\section{Conclusions and Next Steps}
\label{sec:Conclusion}

The experimental results reported here showed that smaller circuit size (program length) allows for a more efficient mapping to match the limitations of the hardware topology. As a result we were able to map larger problem sets to the actual quantum hardware platforms.  This is a critical advancement in the usage of variational quantum algorithms solving such optimization problems.  Using quantum computers with the methodologies discussed here, it is possible to utilize variational algorithms such as VQE to gate allocation problems with large sizes in the near future.

This effort is forward-looking in the sense that actionable solutions to solve these complex gate-scheduling problems are currently beyond the capability of contemporary Noisy Intermediate-Scale Quantum (NISQ) technology.  However, over the past several years there has been significant advancements both in designing and manufacturing quantum hardware processors with larger numbers of qubits and improved performance characteristics.  Advancements in error mitigation and error correction are also shaping the ability of application developers to formulae more complex problems that can be implemented on quantum computers.  

The technology for building larger quantum computer hardware processors is developing rapidly and possibly at some point in the next 5 to 10 years the computational capability of a quantum computer will overtake the capabilities of largest digital computers that can be manufactured for solving these type of NP hard optimization problems.  The results from this basic research focused on developing and applying state-of-the-art quantum algorithms and programming for constrained QAP problems on today's quantum computing hardware platforms will place Delta Air Lines in a "quantum ready" position to capitalize and exploit quantum computing technology in the future as an alternative to today's digital computational methods.

\section{Author Contributions}
HM, YO and MI wrote the python code and ran the simulations and quantum computations on Toronto.  RS and JH organized the test gate-scheduling database and wrote the introduction and classical optimization sections of the paper. PD, HM and MI wrote the quantum optimization, results, conclusions and appendix sections. PD did the overall organization and integration of the material in the paper and PD and MS did the final editing of the document.


%
\begin{APPENDIX}{Superconducting Transmon Quantum Computing Architecture}
\label{AP_QC_hardware}
In this section, we briefly describe the architecture of the \textit{IBMQ Toronto}, a 27-qubit IBM superconducting transmon quantum computing hardware platform used in our experiments. Fig. \ref{IBMQ_Toronto} shows the general architecture and qubit connectivity of the machine and the color coding map that indicates the expected error for each individual qubit and 2-qubit connection (CNOT error). A key aspect about IBM quantum hardware is that each qubit and qubit connection has different error rates. These rates, and other system parameters such as qubit coherence times are not static and change with time. IBM performs periodic calibrations on their quantum systems and the system properties update once this calibration sequence is complete. The properties play a critical role in quantum circuit execution, as they are usually utilized in compiling and optimizing the quantum program. 

The IBM machines are accessed via a cloud portal where users can write quantum computing programs in python using jupyter notebooks. A jupyter notebook can then be submitted either to a quantum computing simulator or one of several different hardware backends onto the actual quantum computing hardware platforms.  The daily measured and published error rates are important factors and considerations when loading programs (circuits) representing quantum algorithms onto a quantum computing hardware platform.   

Programs loaded onto a quantum computer require that each computation in the program be executed on adjacent qubits. The actual physical layout of the qubits shows that there is limited connectivity between each qubit. For large programs it is not possible to place each computation on the qubits so that they are adjacent throughout the execution of the entire program. This limited connectivity of the qubits on the machine requires that the information for the calculation that is contained on distant qubits must to be moved until the two qubits are adjacent.  This is called a \textit{SWAP} operation.  This has a major implication on a quantum program's length that can be executed on a quantum computer before the machine loses its quantum mechanical wavefunction coherence depth in time and stops functioning.  A key goal of compiling a quantum program is to adequately map it to the subset of qubits and connections with the best coherence, error parameters and a connectivity that guarantees the minimum amount of extra computation (SWAP gates) so that it results in the best and longest execution of the program. 

 \end{APPENDIX}
%
%


\bibliographystyle{informs2014} 
\bibliography{bibliography} 



\end{document}